\documentclass[twocolumn]{aastex61}

\newcommand\aastex{AAS\TeX}

\newcommand{\kms}{\hbox{km\,s$^{-1}$}}
\usepackage{hyperref}
\usepackage{natbib}
\received{Jul 3 2017}
\revised{Nov 9 2017}
\accepted{Feb 1 2018}
\submitjournal{PASP}

\shorttitle{\aastex\ A focus on L dwarfs with trigonometric parallaxes}
\shortauthors{Wang et al.}

\begin{document}

\title{A focus on L dwarfs with trigonometric parallaxes}

\correspondingauthor{Youfen Wang}
\email{yfwang@bao.ac.cn}

\author{Y. Wang}
\affiliation{National Astronomical Observatories of China, CAS,
              20A Datun Road Chaoyang District, Beijing, China }
\author{R.L. Smart}
\affiliation{Istituto Nazionale di Astrofisica,
 Osservatorio Astrofisico di Torino,
 Strada Osservatorio 20, 10025 Pino Torinese, Italy}
\affiliation{School of Physics, Astronomy and Mathematics,
 University of Hertfordshire, College Lane, Hatfield AL10 9AB, UK}
\author{Z. Shao}
\affiliation{Shanghai Astronomical Observatory, CAS,
 80 Nandan Road, Shanghai 200030, China}
 \affiliation{Key Lab for Astrophysics, Shanghai 200234, China}
\author{H.R.A Jones}
\affiliation{School of Physics, Astronomy and Mathematics,
 University of Hertfordshire, College Lane, Hatfield AL10 9AB, UK}
\author{F. Marocco}
\affiliation{School of Physics, Astronomy and Mathematics, 
University of Hertfordshire, College Lane, Hatfield AL10 9AB, UK}
\author{A. Luo}
\affiliation{National Astronomical Observatories of China, CAS,
              20A Datun Road Chaoyang District, Beijing, China }
\author{A. Burgasser}
\affiliation{Center for Astrophysics and Space Science,
 University of California San Diego, La Jolla, CA 92093, USA}
\author{J. Zhong}
\affiliation{Shanghai Astronomical Observatory, CAS,
 80 Nandan Road, Shanghai 200030, China}
\author{B. Du}
\affiliation{National Astronomical Observatories of China, CAS,
              20A Datun Road Chaoyang District, Beijing, China }



\begin{abstract}

We report new parallax measurements for ten L and early T type dwarfs, five of
which have no previous published values, using observations over 3 years at
the robotic Liverpool Telescope. The resulting parallaxes and proper motions
have median errors of 2\,mas and 1.5\,mas/year respectively. Their space
motions indicate they are all Galactic disk members. We combined this sample
with other objects with astrometry from the Liverpool Telescope and with
published literature astrometry to construct a sample of 260 L and early T
type dwarfs with measured parallaxes, designated the Astrometry Sample.  We
study the kinematics of the Astrometry Sample, and derived a solar motion of
$(U,V,W)_{\bigodot} = (7.9\pm1.7,13.2\pm1.2,7.2\pm1.0)$\,\kms~ with respect to
the local standard of rest, in agreement with recent literature. We derive a
kinematic age of 1.5-1.7\,Gyr for the Astrometry Sample assuming the age
increases monotonically with the total velocity for a given disk sample. This
kinematic age is less than half literature values for other low mass dwarf
samples.  We believe this difference arises for two reasons (1) the sample is
mainly composed of mid to late L dwarfs which are expected to be relatively
young and (2) the requirement that objects have a measured parallax biases the
sample to the brighter examples which tend to be younger.

\end{abstract}




\section{Introduction} \label{sec:intro}

Objects classified as L and T spectral types are predominately brown dwarfs or
sub-stellar objects with masses that cover the range from the most massive
planets to the least massive stars. Since the discovery of the first examples
\citep{1988Natur.336..656B, 1995Natur.378..463N} there have been over 2000
identified primarily in the large optical and near-infrared (NIR) sky surveys
\citep[e.g. ][]{2000AJ....120..447K, 2004AJ....127.3553K, 2008MNRAS.390..304P}
and more recently in the mid-infrared \citep{2011ApJS..197...19K}.  They are
an important component of the Galaxy that can be used to study the atmospheres
of hot Jupiter-like planets \citep[e.g. ][]{2013MmSAI..84..955F}, to explore
the low mass end of the initial mass function
\citep[e.g. ][]{2012ApJ...753..156K, 2013MNRAS.433..457B,
  2015MNRAS.449.3651M}, and, given their long life time and ubiquity, will be
excellent for studying the evolution of our galaxy and its components
\citep[e.g. ][]{2009IAUS..258..317B}.

Distance is a critical parameter in understanding these objects. A distance is
required to derive the absolute magnitude and, hence, energy output. A
model-independent parallax can be used to constrain radius or temperature,
and, aid in the exploration of relations between other parameters such as
mass, surface gravity, age, and metallicity. To precisely measure a parallax
observational sequences covering several years on stable imaging systems are
required and less than two hundred and fifty of the currently known L and T
dwarfs have measured parallaxes. In this paper we report new parallaxes of ten
L and early T dwarfs and then combine them with all published parallaxes of L0
to T2 dwarfs and examine a number of relations e.g. SpT-absolute magnitude
diagrams and space motions.

This paper is divided into five sections. First, in Section 2 we report the
parallax measurements for our ten targets and use their parameters to indicate
which galactic population they pertain to. In Section 3 we study the spectral
type and absolute magnitude relations. In Section 4 we study the kinematic
signature of the Astrometry Sample. Finally we summarize the paper in Section
5.

\section{Parallax measurements }\label{sec:measurements}

The parallax measurements in this paper were made as part of the program
described in \citet[][ hereafter WJS14]{2014PASP..126...15W}; here we briefly
summarize the observations and data reduction procedures, the reader is
referred to that paper for more details. The observations were made on the
2\,m robotic Liverpool Telescope\footnote{\url{http://telescope.livjm.ac.uk/}}
(hereafter LT).  The LT, an Alt-Az mounted telescope with Ritchey-Chr\'etien
Cassegrain optics, is a totally robotic telescope located at the Observatorio
del Roque de Los Muchachos on the Canary island of La Palma in Spain and
operated by the Liverpool John Moores University in the United Kingdom. We
used the SDSS-$z$ band filter \citep[hereafter simply $z_{AB}$;
][]{2000AJ....120.1579Y} and the RATCam CCD which is an optically sensitive
2048$\times$2048 pixel CCD camera with a pixel scale of 0.1395 {\rm 
arcsecond/pixel} providing a total field of view of 4.6 {\rm arcmin}.

\subsection{Target Selection} 

The targets were selected from the literature with the following criteria: at
a declination visible to the LT, a SDSS $z_{AB}$ magnitude brighter than 18 and
no published trigonometric parallax in 2004.  From this list of objects those
with the smallest photometric distance were preferred. Here we report on ten
objects that have enough observations to provide reliable parallaxes.

In Table~\ref{tb1}, we list the ten targets presented in this contribution and
the five targets from WJS14 with their discovery designation, a short name,
SDSS $z_{AB}$ magnitude, NIR spectral type (hereafter SpT$_{NIR}$), binary
separations from the literature if the object is a known binary system, and,
any published astrometry. SpT$_{NIR}$ is the near-infrared spectral type found
from NIR spectra in the SpeX-Prism spectra library
\footnote{\url{http://pono.ucsd.edu/~adam/browndwarfs/spexprism}}
\citep{2014ASInC..11....7B} comparing standards in the 0.9-1.4\,$\micron$
region \citep{2010ApJS..190..100K}.  The five objects below the solid line in
Table \ref{tb1} are from WJS14, which used the the same methodology as presented
here. These five objects combined with the
ten targets with homogeneous astrometry we designate the LT Sample.

\begin{table*}[h]
\begin{center} 
 \caption{Target list/LT sample with their SDSS $z_{AB}$ magnitude, NIR spectral
   type and published astrometry information. The five objects below the solid line are from WJS14.}
\begin{tabular}{lllllllr}
\hline
2MASS Designation &Short Name  & $z_{AB}$   &SpT$_{NIR}$ & $Binary$&~~v$_r$&Literature $\pi$, $\mu_{\alpha}cos\delta$, $\mu_{\delta}$ \\
                  &            & (mag)     &           & $sep.~(")$  & (\kms) &    (mas, mas/yr, mas/yr) \\ \hline   
J04234858-0414035 & 2M0423-0414 & 17.29 &      T0 &0.61$^1$ &28.0$\pm$2.0$^{5}$&73.3$\pm$1.4, -325.3$\pm$1.0, 93.1$\pm$0.9$^7$\\  
J07171626+5705430 & 2M0717+5705 & {\it 17.23} &L3 &-&-16.3$\pm$0.2$^{6}$&..., -17.99$\pm$17.87, 67.17$\pm$15.13$^8$    \\      
J07584037+3247245 & 2M0758+3247 & 17.96 &      T2 &-&-&..., -204.23$\pm$18.01, -316.21$\pm$12.42 $^8$\\                         
J08575849+5708514 & 2M0857+5708 & 17.74 &      L8 &-&-&..., -413.61$\pm$20.52, -353.43$\pm$16.85$^8$\\ 
J10170754+1308398 & 2M1017+1308 & 16.74 &      L2 &0.10$^{2}$&-& 30.0$\pm$ 1.6, 44.1$\pm$0.7, -114.3$\pm$0.6$^7$\\   
J11040127+1959217 & 2M1104+1959 & 17.21 &      L5 &-&-&..., 74.8$\pm$14.7, 138.7$\pm$20.3$^{9}$ \\           
J12392727+5515371 & 2M1239+5515 & 17.52 &      L6 &0.21$^{3}$&-&42.4$\pm$1.7, 125.2$\pm$1.1, 0.04$\pm$1.1$^7$\\
J13004255+1912354 & 2M1300+1912 & 15.14 &      L1 &-&-17.6$\pm$0.2$^{6}$&70.4$\pm$2.5, -793.0$\pm$10.0, -1231.0$\pm$10.0$^{10}$\\         
J15150083+4847416 & 2M1515+4847 & 16.74 &      L5 &-&-30.0$\pm$0.1$^{6}$&..., -949.9$\pm$21.3, 1471.5$\pm$21.4$^{9}$\\       
J20282035+0052265 & 2M2028+0052 & 16.98 &      L2 &0.05$^{4}$&- &33.25$\pm$1.32, 96.50$\pm$0.93, -6.05$\pm$2.04$^{11}$\\ \hline
J01410321+1804502 & 2M0141+1804 & 16.34 &      L2 &-&24.7$\pm$0.1$^{6}$&44.1$\pm$2.1, 405.2$\pm$1.1, -48.7$\pm$0.9$^{12}$\\
J17171408+6526221 & SD1717+6526 & 17.79 &      L7 &-&-&57.1$\pm$3.5, 105.2$\pm$1.0, -109.3$\pm$0.6$^{12}$\\
J18071593+5015316 & 2M1807+5015 & {\it 15.43} &L1 &-&-0.4$\pm$0.5$^{6}$&77.3$\pm$1.5, 27.2$\pm$1.0, -130.2$\pm$1.5$^{12}$\\
J22380742+4353179 & 2M2238+4353 & {\it 16.42} &L1 &-&-&54.1$\pm$1.6, 324.3$\pm$0.5, -121.0$\pm$0.4$^{12}$\\ 
J22425317+2542573 & 2M2242+2542 & 17.49 &      L2 &-&-&48.0$\pm$2.8, 382.0$\pm$0.9, -64.6$\pm$0.7$^{12}$\\ \hline          
\end{tabular}\\
\medskip
 \label{tb1}
\end{center}
Note: All magnitudes are measured SDSS~DR10 $z_{AB}$ magnitudes except the
three objects in italics which are estimated from their $J$ and $K$ magnitudes.
The SpT$_{NIR}$ is estimated in this work following the 
\citet{2010ApJS..190..100K} method using SpeX-Prism spectra. The $Binary~sep.$
indicates the angular separation in arcsecond for known binary systems. The
last column listed literature absolute parallax and proper motions when
available. The five objects below the line are from WJS14. \\ References:
$^1$\cite{2006ApJ...637.1067B}, $^{2}$\citet{2003AJ....126.1526B}, 
$^{3}$\citet{2003AJ....125.3302G}, $^{4}$\citet{2013ApJ...767..110P}, $^{5}$\citet{2015ApJ...808...12P}, 
$^{6}$\citet{2010ApJ...723..684B}, $^{7}$\citet{2012ApJS..201...19D}, 
$^{8}$\citet{2008MNRAS.390.1517C}, $^{9}$\citet{2008MNRAS.384.1399J}, 
$^{10}$\citet{2016ApJS..225...10F}, $^{11}$\citet{2016AJ....152...24W}, 
$^{12}$\citet{2014PASP..126...15W}. \\
\end{table*}

\subsection{Observations} 

All observations were obtained within 30\, minutes of the meridian to minimize
Differential Color Refraction
\citep{1992AJ....103..638M,2002PASP..114.1070S}.  This is the small varying
positional displacement of objects due to their different colors and 
the variation of the atmosphere refractive index with wavelength. Observing at
small hour angles minimizes the part of the refraction that varies as a function
of the amount of atmosphere traversed. In each observation we took three
exposures of 160\,s to allow for robust removal of cosmic rays and to minimize
random errors. This combination of exposures nominally provides a
signal-to-noise of better than 50 on these targets.

\subsection{Data reduction} 
The bias subtraction, trimming of the over-scan regions, dark subtraction and
flat fielding are carried out via the standard LT pipeline
\citep{2004SPIE.5489..679S}.  Images in the z band display prominent fringes
caused by thin-film interference \citep{2008A&A...488..533B}. These fringes
can have a significant impact on the astrometry of our targets since the targets
are relatively faint. The LT web site provides biannual fringe maps
which we used to remove the fringes using IRAF's {\it rmfringe}. We derived
the x and y positions using a maximum likelihood barycenter centroid as
implemented in the {\it imcore} software of the Cambridge Astronomy Survey
Unit (hereafter
CASU\footnote{\url{http://casu.ast.cam.ac.uk/surveys-projects}}).
We compared successive observations of the same field and find the centroid
precision is approximately 11\,mas for bright objects in both x and y
coordinates (WJS14).

\subsection{Parallax determination and comparison}
We derived the parallaxes and proper motions using the methods adopted in the
Torino Observatory Parallax Program \citep[][]
{2003A&A...404..317S,2007A&A...464..787S}, using the x,y coordinates
determined from the CASU {\it imcore} software. The Torino pipeline selects
the frames and reference stars in an unsupervised fashion with user supplied
parameters to vary the minimum number of common reference stars and the
outlier rejection criteria. A base frame is selected in the middle of the
sequence with a high number of stars. This base frame is transferred to a
standard coordinate system using the Sloan Digital Sky Survey
\citep{2000AJ....120.1579Y} as a reference catalog except for 2M0717+5705
where we used the Two Micron Sky Survey \citep{2006AJ....131.1163S}.

The other observations of each target are translated to the base frame
standard coordinate system using all common stars via a linear
transformation. Once we have all observations in the base frame system we fit
the observations of the target with a position offset, parallax and proper
motion in each coordinate. The best relative parallax is found from a weighted
mean of the estimates in each coordinate. The correction from relative to
absolute parallax is calculated using the Galaxy model of
\citet{1998A&A...330..910M} as described in \citet{2003A&A...404..317S}. We
estimate the error on this correction to be around 30\% or 0.4-0.6\,mas for
these fields which is negligible compared to the formal error of the
parallaxes.

In Table~\ref{tb2} we list the parallax and proper motions of ten L/T dwarfs.
The motion and corresponding fit over the observed period for all targets are
shown in the appendix Fig.~\ref{elis}.  In the last column of Table~\ref{tb1},
published parallax and proper motion results are shown. We found that five of
our targets have literature parallaxes for which our values are all consistent
to within two $\sigma$ except for the target 2M2028+0052 which differs by
three times the combined $\sigma$ from the value in
\citet{2016AJ....152...24W}.  2M2028+0052 is a known binary with almost equal
mass and magnitude components \citep{2013ApJ...767..110P} while both the
Weinberger and the LT solutions assume it is single. The LT solution has more
epochs, 17 vs 4, and a longer baseline, 3.74 vs 2.0\,yrs, so we expect the
results presented here to be more robust.

\begin{table*}
\begin{center} 
 \caption{Parallax and proper motions for our ten targets.}
\begin{tabular}{lrrrrrrr}\hline
 Short Name & $N_{\rm obs}$,$N_{\rm ref}$ & $\Delta$t & $\pi$ & COR & $\mu_{\alpha}cos\delta$ & $\mu_{\delta}$&v$_{tan}$\\ 
  & & { (yr)}&{ (mas)}&{ (mas)}&{ (mas)}&{ (mas)}&{ (mas)}\\ \hline
 2M0423-0414 &50,13& 4.33   &66.3$\pm$3.7  & 1.5  &-325.6$\pm$1.8  &  83.1$\pm$1.5& 24.0$\pm$1.3 \\
 2M0717+5705 &48,27& 4.22   &46.4$\pm$2.3  & 1.4  &-18.0$\pm$1.6   &  54.3$\pm$1.3&5.9$\pm$0.3 \\
 2M0758+3247 &62,16& 4.18   &106.9$\pm$4.6 & 1.5  &-230.7$\pm$1.7  & -327.7$\pm$2.1&17.9$\pm$0.8 \\
 2M0857+5708 &71,9 & 4.15   &98.0$\pm$2.6  & 2.0  &-400.0$\pm$1.7  & -374.9$\pm$1.7&26.5$\pm$0.8\\
 2M1017+1308 &55,5 & 4.10   & 32.3$\pm$2.8 & 2.5  & 61.0$\pm$1.4   & -116.3$\pm$1.5&19.3$\pm$1.7\\
 2M1104+1959 &66,5 & 4.08   & 66.2$\pm$1.9 & 2.2  & 55.9$\pm$0.7   & 126.6$\pm$0.7&9.9$\pm$0.3\\
 2M1239+5515 &43,6 & 4.04   &45.0$\pm$2.1  & 2.0  &131.7$\pm$2.0   &  -2.6$\pm$1.4&13.9$\pm$0.7 \\
 2M1300+1912 &42,10& 3.48   &76.4$\pm$1.8  & 2.0  &-789.0$\pm$1.1  &-1237.2$\pm$1.0&91.1$\pm$2.1 \\
 2M1515+4847 &43,6 & 3.46   &123.8$\pm$5.0 & 1.7  &-930.4$\pm$4.1  & 1469.3$\pm$2.2&66.8$\pm$2.5\\
 2M2028+0052 &54,79& 3.74   & 39.1$\pm$1.6 & 1.2  & 96.9$\pm$0.8   &  -9.0 $\pm$0.8&11.8$\pm$0.5\\ \hline
\end{tabular}\\
 \label{tb2}
\end{center} 
 { Note:} The columns denote the object name, number of observations and
 number of reference objects ($N_{\rm obs}$,$N_{\rm ref}$), total time span
 for observations ($\Delta$t), absolute parallax ($\pi$ ), correction from
 relative to absolute parallax (COR), proper motions ($\mu_{\alpha}cos\delta$
 \& $\mu_{\delta}$) and tangential velocity($v_{tan}$). 
\end{table*}

\subsection{Galactic population membership}

Kinematic information can be used as an indication of Galactic population
membership.  Eight of our ten targets have low tangential velocities of $<$
25\,\kms~ while 2M1515+4847 has v$_{tan}$=66$\pm$2.5\,\kms~ and 2M1300+1912 has
v$_{tan}$=91.1$\pm$2.1\,\kms. The galaxy model of v$_{tan}$ shown in Fig.~31
of \citet{2012ApJS..201...19D} implies that all these targets are thin disk,
although 2M1300+1912 may be thick disk.

\begin{figure}[htbp]
\centering
\includegraphics[width=0.33\textwidth]{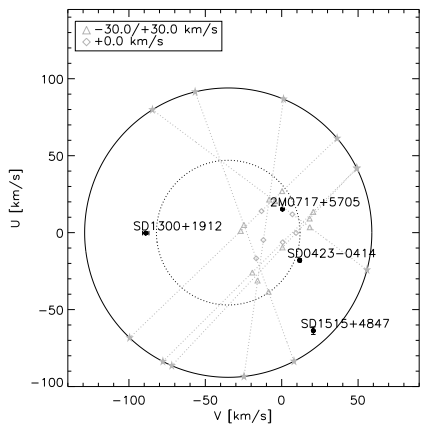}
\caption{U versus-V Galactic velocities of our ten targets. The dotted and
  solid circles are 1$\sigma$ and 2$\sigma$ velocity ellipsoids for disk stars
  with the center at (-45,0)\,\kms~ and radii of 47\,\kms~ and 94\,\kms~
  \citep{2001ApJ...559..942R,2001Sci...292..698O}.  The labeled solid dots
  with error bars indicate the four targets with measured radial velocities.
  The six grey dashed lines indicate six targets without measured radial
  velocity, the lines also indicate the U and V velocities distribution for
  each objects when assuming each of them have Gaussian radial velocity
  distribution centered in 0 and 1$\sigma$ of 30\,\kms. }
\label{uvplot}
\end{figure}

UVW space velocities can also be used to indicate Galaxy
population. Determination of U and V requires a measurement of parallax,
proper motion and radial velocity. Four of our ten targets have radial
velocity measurements and these are given in Table~\ref{tb1}.  For the
remaining targets we calculate their U and V velocities assuming a Gaussian
distribution of radial velocities centered on zero with a $\sigma=$30\,\kms,
as seen for M dwarfs radial velocity (WSJ14). Fig.~\ref{uvplot} shows the UV
velocities for our ten targets, as well as the one and two sigma disk stars'
velocity ellipsoids \citep{2001ApJ...559..942R,2001Sci...292..698O}. Targets
beyond the two sigma ellipsoid with $[U^2 + (V + 35)^2]^{1/2} > 94$\,\kms~ are
likely to be halo members \citep{2001Sci...292..698O}. In this view all the
targets appear to be likely thin disk members although SDSS 1515+4847 lies
near the two sigma ellipsoid and may be a thick disk object.

We can also calculate a probability of these objects being Galactic disk
members.  Adopting a Gaussian distribution for radial velocity, these objects
have U and V velocities that trace a straight line on a U-V plot as shown in
Fig.~\ref{uvplot}. Integrating within the 2$\sigma$ circle, we derived the
probability of them being Galactic disk members. The probability is very high
$\sim$100\% for all of our targets indicating that they are
Galactic disk members.

\section{Study of Spectral type versus absolute magnitude diagrams}\label{sec:sptvsmag}

We now examine the Spectral Type versus absolute Magnitude
relations (hereafter SpT-Mag) in the 2MASS and WISE magnitude systems.

\subsection{The Astrometry Sample and the LT Sample}

To obtain a larger, statistically significant, sample we combine all brown
dwarfs with spectral types between L0 to T2 and published trigonometric
parallax measurements. This sample is a combination of the
\citet{2012ApJS..201...19D} online compendium, 5 targets from WJS14 and the 10
targets presented in this contribution resulting in 260 objects (239 L and 21
T dwarfs), designated the Astrometry Sample. In the online table we list
object name, position, optical and NIR spectral types, 2MASS magnitudes,
object flag, WISE magnitudes, trigonometric distance, tangential velocity,
radial velocity, $U V W$ space motions and trim status in Section 5 for our
Astrometry Sample. We will use the Astrometry Sample when plotting the SpT-Mag
diagrams in the next subsection and in the Section 4.

The LT Sample (Table~\ref{tb1}) is a sub sample of the Astrometry Sample that
contains the ten targets with new astrometry presented here and the five
targets presented in WJS14. This LT sample is considered separately because
it has homogeneous photometry, spectroscopy and astrometry.

%

\subsection{SpT-Mag diagrams}

\begin{figure*}[]
\centering
\includegraphics[width=3.in]{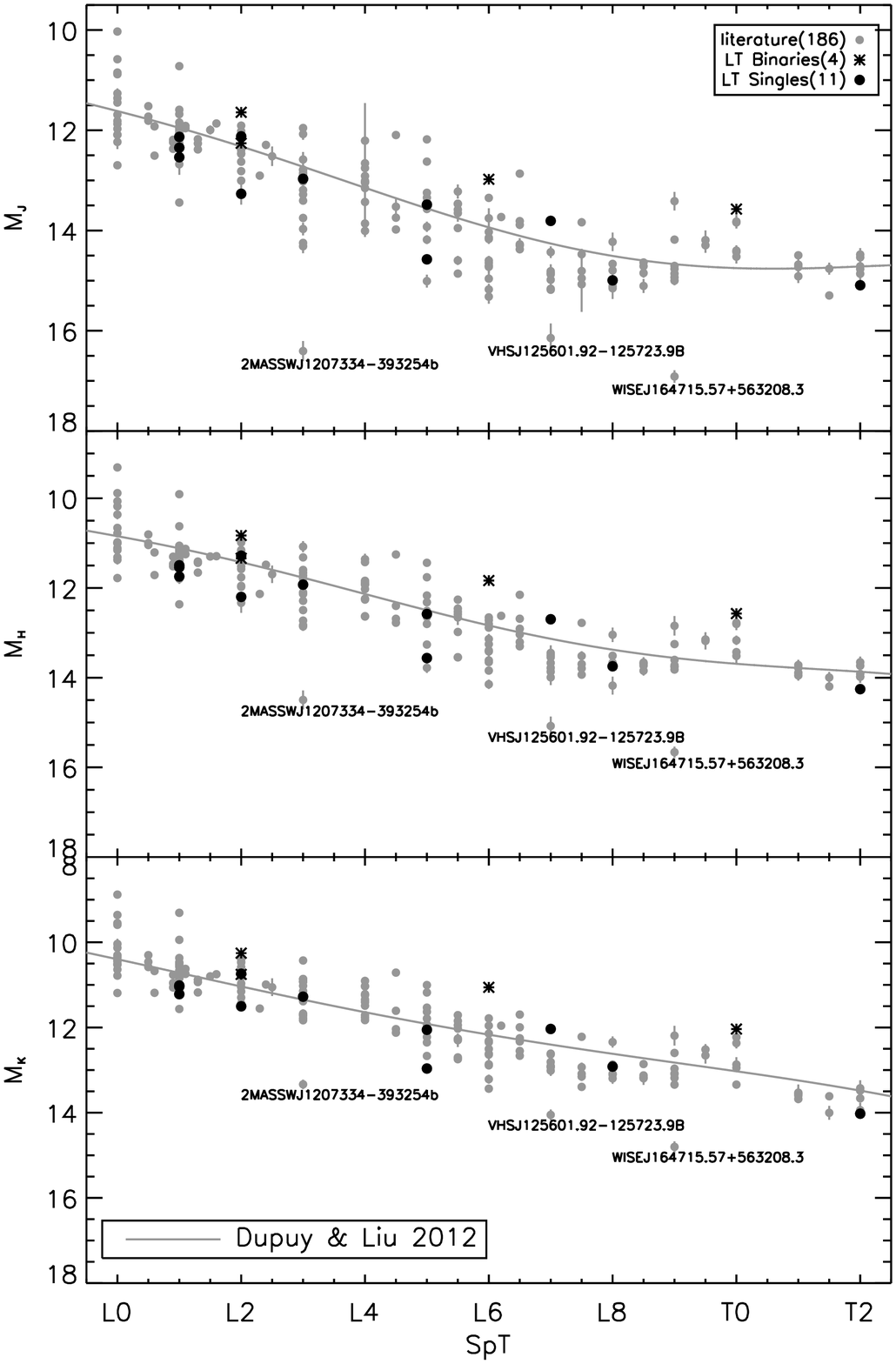}
\includegraphics[width=3.in]{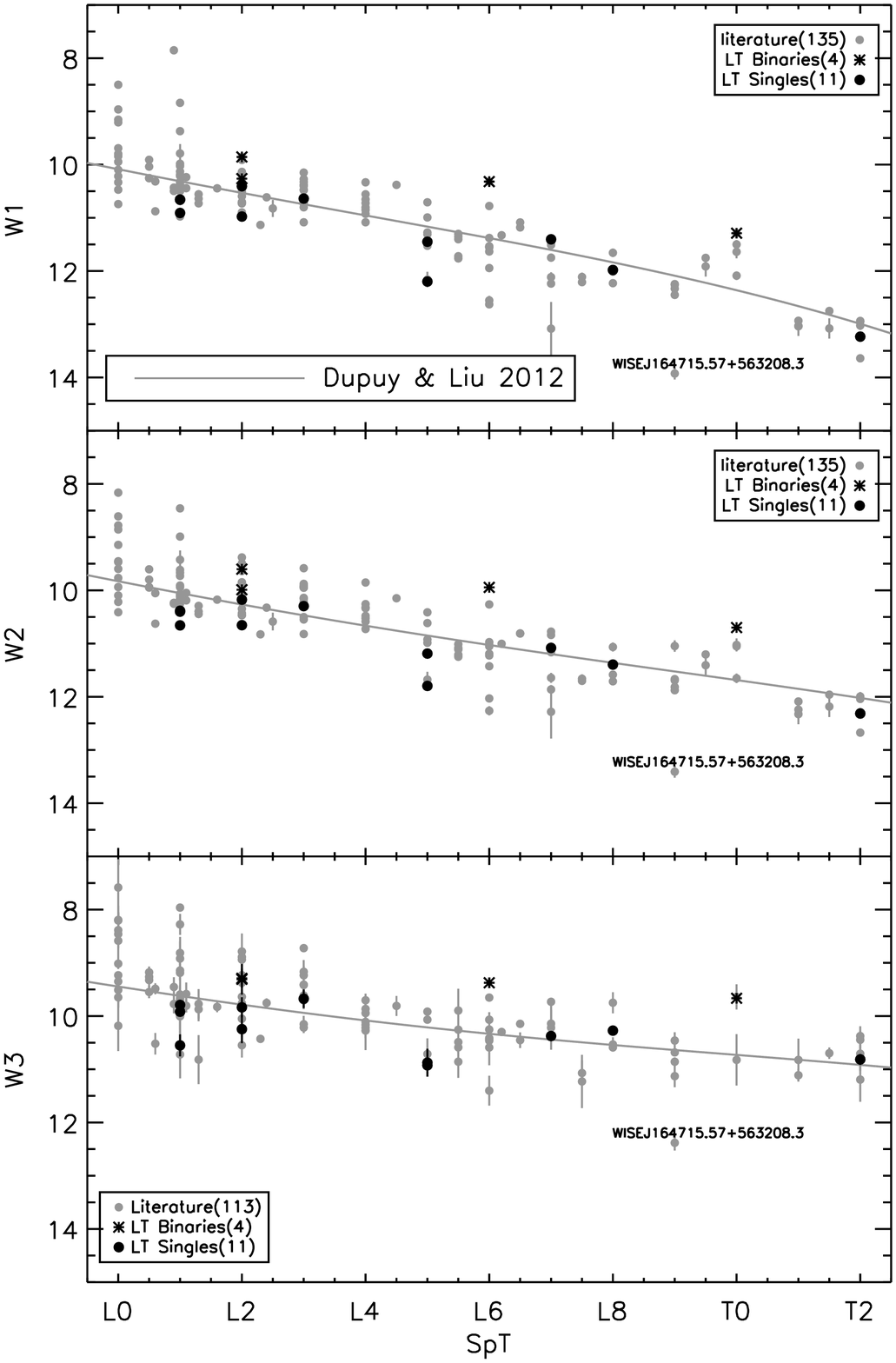}
\caption{Spectral type versus absolute magnitudes in 2MASS and WISE system.
The grey solid dots are the literature objects with parallax measurements.
The grey solid fitting line is from \citet{2012ApJS..201...19D}. The LT sample are
 indicated as black symbols, with black solid circles for singles, small black 
asterisks for known binaries. The outliers are labeled.}
\label{spt_mag_2MASS}
\end{figure*}

In Fig.~\ref{spt_mag_2MASS}, we plot the SpT-Mag diagrams in the 2MASS and
WISE systems respectively. The grey solid circles indicate the objects from
Astrometry Sample, the black symbols indicate the LT Sample. The small black
asterisks indicate four LT sample targets that are known binaries. The solid
grey line is the relation from \citet{2012ApJS..201...19D} using a sixth order
polynomial fit.

There are 200 objects with valid $2MASS ~ J H K_s$ magnitudes plotted in the  
left panel of Fig.~\ref{spt_mag_2MASS}. We have labeled three under-luminous outliers in 
each 2MASS band. Both 2MASSWJ1207334-393254B \citep{2013ApJ...772...79A} and 
VHSJ125601.92-125723.9B \citep{2015ApJ...804...96G}  
have low surface gravity and young age. While WISEJ164715.57+563208.3 has a very red 
near-infrared color which cannot be attributed to low gravity \citep{2011ApJS..197...19K}.

For the plots in WISE absolute magnitudes to maximize the numbers of objects
we selected on W1 and W2 separately from W3, all objects with values were
considered valid regardless of the flags and error estimates. This resulted in
151 objects in the W1 and W2 bands and 129 objects in the W3 band plotted in
the right panel of Fig.~\ref{spt_mag_2MASS}.  The grey solid line shows the
polynomial fit presented in \citet{2012ApJS..201...19D}.  

The outliers 2MASSWJ1207334-393254B and VHSJ12\\5601.92-125723.9B do not have
valid WISE magnitudes as they are both blended with bright
stars. WISEJ164715.57+563208.3 remains an outlier in the WISE 
bands. 

The consistency of our data with the published data in the SpT-Mag diagrams
is a confirmation that our parallax measurements are reasonable. In the 
SpT-Mag diagram the unresolved binaries stand out as over-luminous objects,
up to 0.75 magnitudes for equal mass binaries.  There are four of our ten targets located above the
fitting line in both panels of Fig.~\ref{spt_mag_2MASS} that are the known
binaries 2M0423-0414 \citep{2006ApJS..166..585B}, 2M1017+1308
\citep{2003AJ....126.1526B}, 2M1239+5515 \citep{2003AJ....125.3302G} and
2M2028+0052 \citep{2013ApJ...767..110P} with separations listed in
Table~\ref{tb1}. 

\section{Kinematic analysis}

In our Astrometry Sample we have 260 objects consisting of 239 L and 21 T
dwarfs. There are 22 known binary systems in our sample each of which we treat
as just one tracer for our kinematical analysis, hence we have 238 tracers for
consideration. These tracers are distributed within $\sim$100\,pc in distance
with a median value of 21\,pc and represent the very close Solar neighborhood.
In this sample, 70 objects have radial velocity ($v_r$) measurements: 41 from
\cite{2010ApJ...723..684B} with uncertainties usually less than 0.2\,\kms, and
29 are gathered from various sources
\citep{2015ApJS..220...18B,2010A&A...512A..37S,2015ApJ...808...12P,
  2000ApJ...538..363B,2009ApJ...705.1416R} with typical uncertainties of
$\sim$2\,\kms.

\subsection{Galactic motions}

\begin{figure*}[ht]
\includegraphics[width=0.33\textwidth,angle=-90]{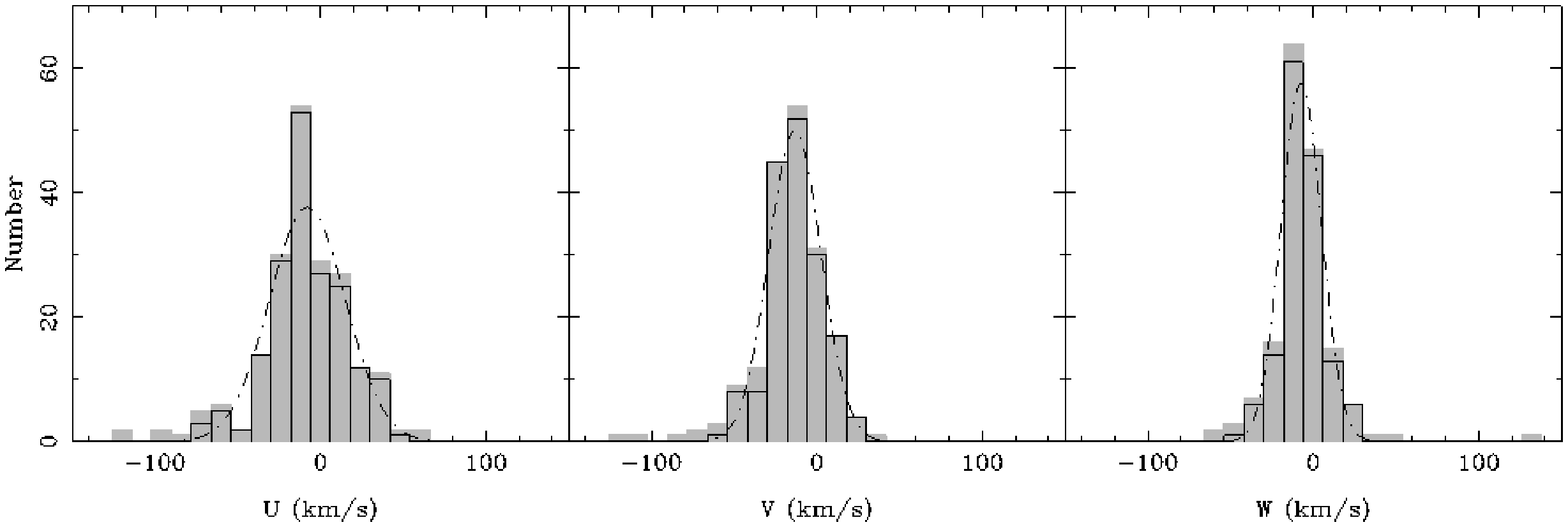}
\caption{Distributions of $U$,$V$ and $W$ velocity components with their best
  fitted Gaussian curves(dot-dashed lines). The solid gray histograms are for whole 
sample and that within the black solid lines are for the trimmed sample. }
\label{Hist_UVW}
\end{figure*}

The majority of the dwarfs in our sample do not have radial velocities.
However we can estimate two out of the three UVW velocities by determining
which motion is most dependent on the unknown radial motion from the target
location and calculating the other two velocities assuming
a radial velocity of zero \citep{2013AJ....145..102L}. We tested this
procedure with simulations and comparing the global parameters of the
sub-sample with known radial velocities and found no evidence for biases.
Together with the 70 objects with measured radial velocities, we find 200, 182
and 164 measurements of U, V and W velocities respectively.  The distributions
of these three velocity components are shown as histograms in
Fig.~\ref{Hist_UVW}.

\subsection{Velocity distributions}

We first examine the velocity distribution properties in the Galactic
reference frame, i.e. the average values and the dispersions of all $U$, $V$
and $W$ components. To exclude high velocity objects we use $3\sigma$ clipping
in all of the $U$, $V$ and $W$ distributions. If an object is rejected from
one of these velocity components, it will also not be used in the other two
components. An iterative process was employed to produce a clean final sample
yielding 21 outliers and 217 tracers.  The trimmed tracer numbers ($n_t$)
for $U$,$V$ and $W$ components are 181, 167 and 145. Their average velocities
($<v>$) and dispersions ($\sigma$) are listed in Table~\ref{table-UVW}.  In
Fig.~\ref{Hist_UVW} we see the trimmed sample dwarfs are well matched to
Gaussian distributions. A Kolmogorov--Smirnov test of the distributions
indicates the untrimmed data is not Gaussian at the 95\% level while the 
trimmed samples are, from which we conclude the cleaning process is required.

\begin{table}{}
\begin{center}
  \caption{Galactic velocity distributions of sample dwarfs }\label{table-UVW}
  \begin{tabular}{ccrrrrr}
  \hline
     & $n$, $n_t$,$n_W$& $<v>$~	  & $\sigma_v$~~~  &  $\sigma_{v(|W|-<>)}$ \\
     &                 & (\kms)     & (\kms)        & (\kms)  \\
  \hline
   $U$ & 200, 181, 111 & $ -7.9\pm1.7$ & $23.0\pm1.3$ & $22.9\pm0.5$ \\
   $V$ & 182, 165, 95 & $-13.2\pm1.2$ & $15.8\pm0.9$ & $18.4\pm0.5$ \\
   $W$ & 164, 147, 147 & $ -7.2\pm1.0$ & $12.2\pm0.7$ & $12.6\pm0.2$ \\
  \hline
{$\sum$}&              &               & $30.5\pm1.7$ &$32.0\pm0.7$ \\
  \hline
  \end{tabular}
\end{center}
{Columns 1 to 6: (1) the space velocity components; 
(2) the derived number of tracers for each space velocity component, the number
 after 3$\sigma$ trimming and number of tracers with velocity of W; (3) mean 
velocity; (4) dispersion;(5) W-weighted velocity dispersion.}
\end{table}

Our velocity data are all heliocentric, so the average value of the sample
reflects the anti-motion of the solar system relative to these dwarfs. Using
our dwarfs for reference we find $(U,V,W)_{\bigodot} =
(7.9\pm1.7,13.2\pm1.2,7.2\pm1.0)$\,\kms. These average values agree
within 2$\sigma$ of recent literature results for the solar motion,
e.g., \cite{2010MNRAS.403.1829S} re-examine the HIPPARCOS data and conclude
that $(U,V,W)_{\bigodot} = (11.10^{+0.69}_{-0.75}, 12.24^{+0.47}_{-0.47},
7.25^{+0.37}_{-0.36})$\,\kms, with additional systematic uncertainties
$\sim$(1, 2, 0.5)\,\kms, and \cite{2015MNRAS.449..162H} derive
$(U,V,W)_{\bigodot} = (7.01\pm0.20,10.13\pm0.12,4.95\pm0.09)$\,\kms,
based on radial velocities from the LAMOST results
\citep{2015arXiv150501570L}. These agreements, especially in the Galactic
rotation direction ($V$), indicate that our dwarf sample is not much different
from the motion of the local standard of rest.

The velocity dispersions of our sample are $(\sigma_{U}, \sigma_{V},
\sigma_{W}) = (23.0\pm1.3, 15.8\pm0.9, 12.2\pm0.7)$\,\kms.  We compare our
values to other velocity dispersions that focus on late M, L and T dwarfs, e.g.,
(30.2, 16.5, 15.8)\,\kms~ from \citet{2007ApJ...666.1205Z}, (22, 28, 17)\,\kms~
from \citet{2009AJ....137....1F}, (25, 23, 20)\,\kms~ from
\citet{2010AJ....139.1808S} and (33.8, 28.0, 16.3)\,\kms~ from
\citet{2010A&A...512A..37S}. Our results are consistent but systematically 
smaller than these literature values. These differences might be due to small
sample sizes, incomplete outlier exclusion or, as we require there to be a
parallax determination, our sample will be biased to brighter, hence younger,
examples.

\subsection{Kinematical age estimation}

Since there are various heating process in the dynamic evolution of the disk, it
was found that there is a monotonic increase of the velocity dispersion with
the mean age of a given stellar population \citep{1977A&A....60..263W}.

We use two methods to find ages from this empirical relation. Firstly, we
employ the velocity-dependent diffusion relationship of
\cite{1977A&A....60..263W}, equation 13 for age $<3$\,Gyr:
\begin{equation}\label{age1977}
    \tilde{\sigma}_{v}(\tau) = (\sigma_0^3 + 1.5\gamma_v \tau)^{1/3},
\end{equation}
where $\tau$ is the statistical age measured in Gyr, $\sigma_0=10$\,\kms~
and $\gamma_v=1.4\times 10^4$ (\kms)$^3$\,Gyr$^{-1}$ and
$\tilde{\sigma}_{v}$ is the total velocity dispersion measured by
$|W|$-weighted velocity dispersion of all three components.
The dispersion results, together with the number of dwarfs with $W$-velocity
($n_W$), are listed in Table~\ref{table-UVW}. For $\tilde{\sigma}_{v} =
32.0\pm0.7$\,\kms, we obtain the age of our sample as $1.5\pm0.1$\,Gyr.

Secondly we follow the development in  \cite{2008gady.book.....B} via the
power-law relation: 
\begin{equation}\label{age2008}
    \sigma_v(\tau) = v_{10} \left( \frac{\tau + \tau_1}{10{\rm Gyr} + \tau_1} \right)^{\beta}
\end{equation}
where $\sigma_v$ is the unweighted total
velocity dispersion, and we used all six best-fit parameter sets of $v_{10}$,
$\tau_1$ and $\beta$ in Table 2 of \cite{2009MNRAS.397.1286A} to provide 
an average age in ($\tau$). With $\sigma_v = 30.5 \pm 1.7$\,\kms, we
find $\tau = 1.7\pm0.3$\,Gyr. The results of both above methods are
consistent. This time scale is roughly 8 times longer than the orbital period
of the Galactic rotation at the solar position, so our sample dwarfs should be
kinematically mixed with the disk, which is also expressed by the velocity
dispersion ratios.

\begin{table}
 \caption{Ages of nearby objects.}
\centering
\begin{tabular}{clcc}
\hline
Ref.  &  Sample  & Method  & Age\,(Gyr)    \\
\hline
R1 & 63 M7-M9.5 $d<20$pc dwarfs      & 1   & 3.1 \\
R2 & 43 L dwarfs                     & 1  & $\sim5.1$ \\
R3 & 16 normal colour late-L dwarfs  & 1,2 & $3.4\sim3.8$ \\
R3 & 28 unusually blue L dwarfs      & 1,2 & $5.5\sim6.5$ \\
\hline
\end{tabular}\\ 
Ref. R1: \cite{2009ApJ...705.1416R}, R2: \cite{2010A&A...512A..37S}, R3:
\cite{2015ApJS..220...18B}. Method 1 from \cite{1977A&A....60..263W}
 and method 2 from \cite{2008gady.book.....B}.
\label{ages}
\end{table}  

In Table~\ref{ages} we list age estimations from literature which use the same
methods as here but with different samples. The estimated kinematic ages for
nearby low mass star and brown dwarfs have a large spread, however, our sample
is found to be significantly younger. We note that our kinematic age
estimation is directly related to the total velocity dispersion. Velocity
dispersion can be affected by the sample size, sample population, and,
peculiar objects with large velocity etc.  In Table~\ref{ages} each sample -
late M dwarfs, small samples, color selected objects - have characteristics
that could produce larger velocity dispersions. Our sample is significantly
bigger than those in Table~\ref{ages} and the outlier rejection is crucial to
remove contamination by thick disk or halo objects. With this in mind we
review the 21 rejected outliers.

Comparing the $J-K_s$ colors of the selected and rejected samples we find the
median value of the 21 outliers is 0.2 mag bluer than that of the normal
dwarfs, though both outliers and the normal dwarf sample have large
dispersions in $J-K_s$ color.  This finding is consistent with the result of
\citet{2015ApJS..220...18B} that unusually blue L-dwarfs have a large velocity
dispersion.  Since our goal is to have a tracer sample and not a
volume-complete sample we believe a rejection of 10\% should not form a
significant bias and hence we are confident of our rejection criteria.

Fig.~8 of \citet{2004ApJS..155..191B} shows that the median age of late M and 
very early L dwarfs, where you still have main sequence objects, is $\sim$4\,Gyr,
 while mid- to late-Ls have a median age of below 2\,Gyr and T dwarfs have a median
age of $\sim$5\,Gyr.  As our sample is predominantly mid- to late-L dwarfs an
age of about $\sim$2\,Gyr is therefore not unexpected. Also considering our
selection criteria that required the targets to have a measured parallax which 
combined with the normal procedures in building parallax target lists means
that our sample will be biased to brighter candidates, which tend to be
younger.

\section{Conclusion}

In this paper we report new parallax measurements for ten L and early T type
dwarfs using the robotic LT telescope, of these, five had no previous distance
determinations. We used the same method as WJS14 adopted for five L dwarfs
using the LT SDSS $z_{AB}$ band data. We study their motions and conclude that
they are probably members of the galactic disk.

The location of our ten targets in the SpT-Mag diagrams have shown the
reliability of our trigonometric parallax measurements. In the 2MASS and WISE
absolute magnitude versus spectral type diagrams we find four LT targets are
over-luminous which are the known binaries 2M0423-0414, 2M1017+1308,
2M1239+5515 and 2M2028+0052.

We have combined our sample with literature L and T dwarfs compiling a list of
260 objects with measured parallaxes, proper motions, radial velocities (for
70 objects) as well as 2MASS and WISE magnitudes which are listed in the
online table. We study the velocity distribution and the kinematic age of this
sample. We derive the solar motion $(U,V,W)_{\bigodot}
=(7.9\pm1.7,13.2\pm1.2,7.2\pm1.0)$\,\kms, which is consistent with recent
literature. The velocity dispersion of our sample is
$(\sigma_{U},\sigma_{V},\sigma_{W})=(23.0\pm1.3,15.8\pm0.9,12.2\pm0.7)$\,\kms.
The kinematical age of our sample is 1.5-1.7\,Gyr, significantly younger than
other estimates for the ages of other samples of late M and L dwarfs. We
believe that this arises because our sample is dominated by mid to late L
dwarfs, and, biased to intrinsically brighter, therefore younger, examples.
 

\acknowledgments 

We wish to thank the referee, Dr Sandy Leggett, for insightful and useful
comments on the original submission of this paper. We thank Leigh Smith from
the University of Hertfordshire for his help in analyzing the image data;
Zhenghong Tang, Yong Yu, and Zhaoxiang Qi from Shanghai Astronomical
Observatory for their helpful discussion on the centroiding precision; and
Yihan Song from National Astronomical Observatories of China for his helpful
discussion on the proper motion transformation. This work is partially funded
by IPERCOOL n.247593 International Research Staff Exchange Scheme and PARSEC
n.236735 International Fellowship within the Marie Curie 7th European
Community Framework Programme.  RLS's research was supported by a visiting
professorship from the Leverhulme Trust (VP1-2015-063). YW and AL acknowledges
NSFC grant NO. 11233004.  ZS acknowledges NSFC grant NO. 11390373. HRAJ and FM
acknowledges support from the UK's Science and Technology Facilities Council
[grant number ST/M001008/1]. JZ acknowledge the NSFC Grant No.11503066 and the
Shanghai Natural Science Foundation (14ZR1446900).  This research has
benefited from the M, L, and T dwarf compendium housed at DwarfArchive.org and
maintained by Chris Gelino, Davy Kirkpatrick, and Adam Burgasser. This research 
has benefited from the SpeX-Prism Spectral Libraries, maintained by Adam 
Burgasser at http://wwww.browndwarfs.org/spexprism. The
Liverpool Telescope is operated on the island of La Palma by Liverpool John
Moores University in the Spanish Observatorio del Roque de los Muchachos of
the Instituto de Astrofisica de Canarias with financial support from the UK
Science and Technology Facilities Council.

\appendix

\section{Figures}
\begin{figure*}
\gridline{\fig{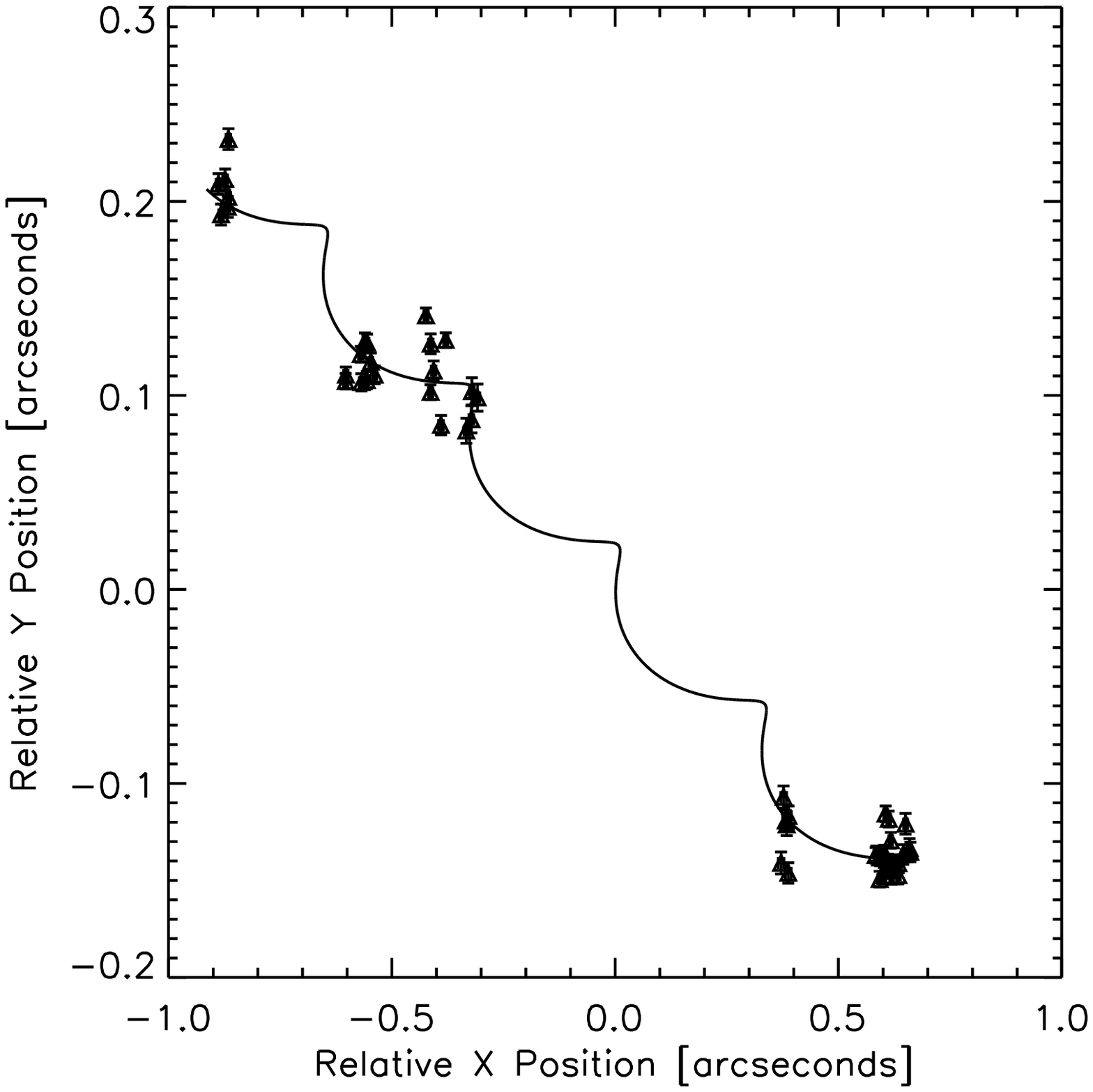}{0.25\textwidth}{2M0423-0414}
          \fig{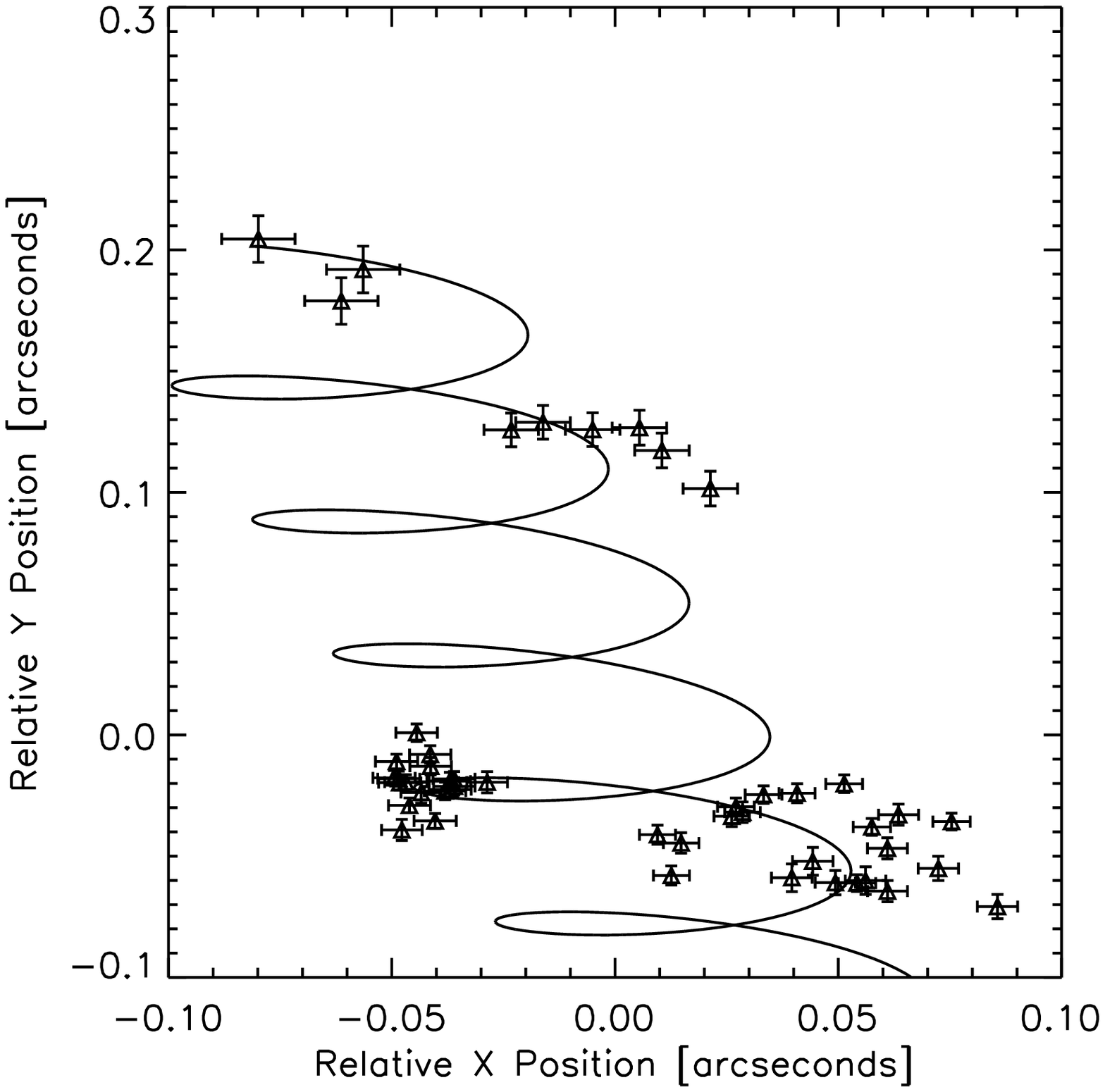}{0.25\textwidth}{2M0717+5705}
          \fig{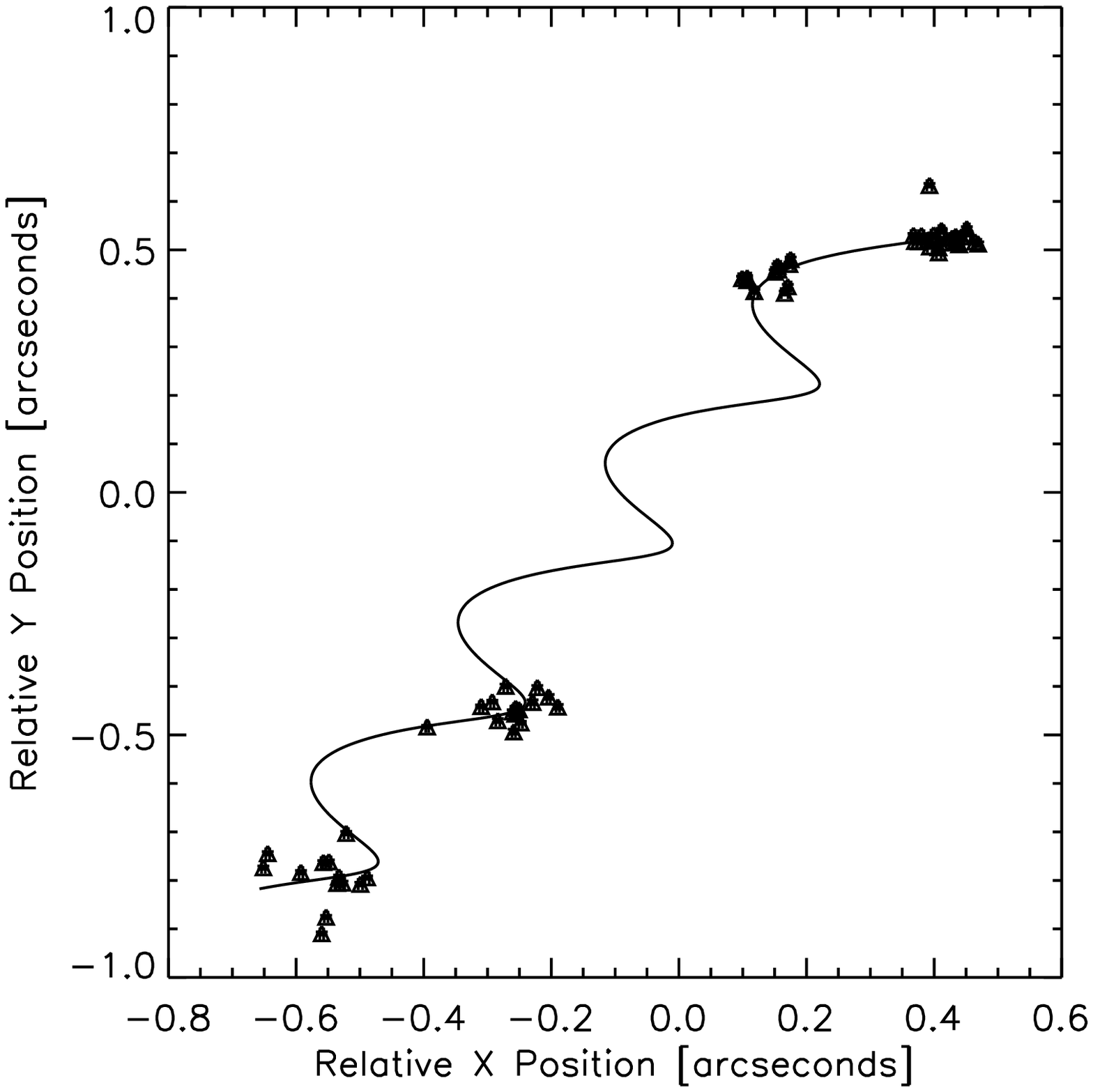}{0.25\textwidth}{2M0758+3247}
          \fig{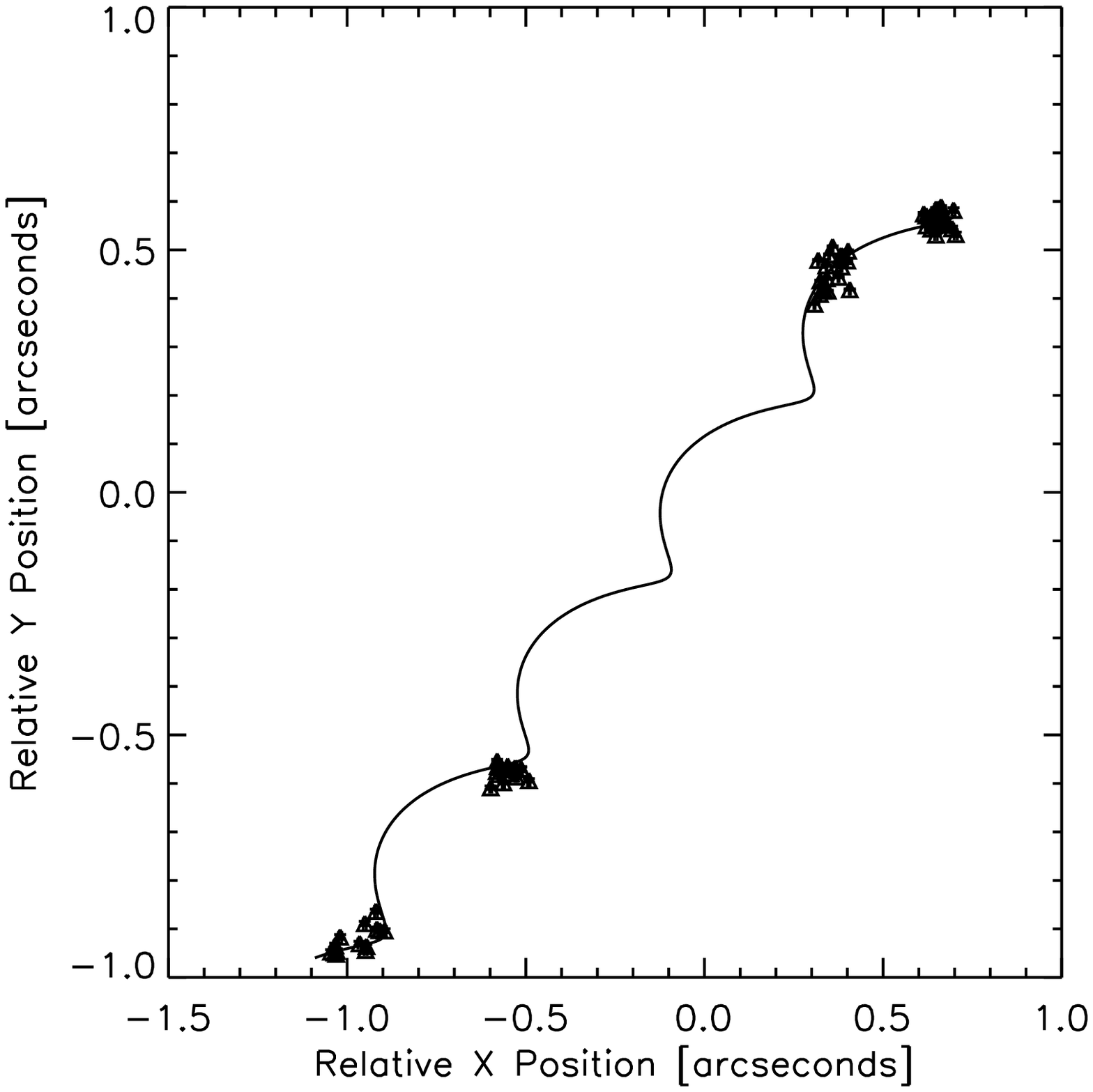}{0.25\textwidth}{2M0857+5708}
          }
\gridline{\fig{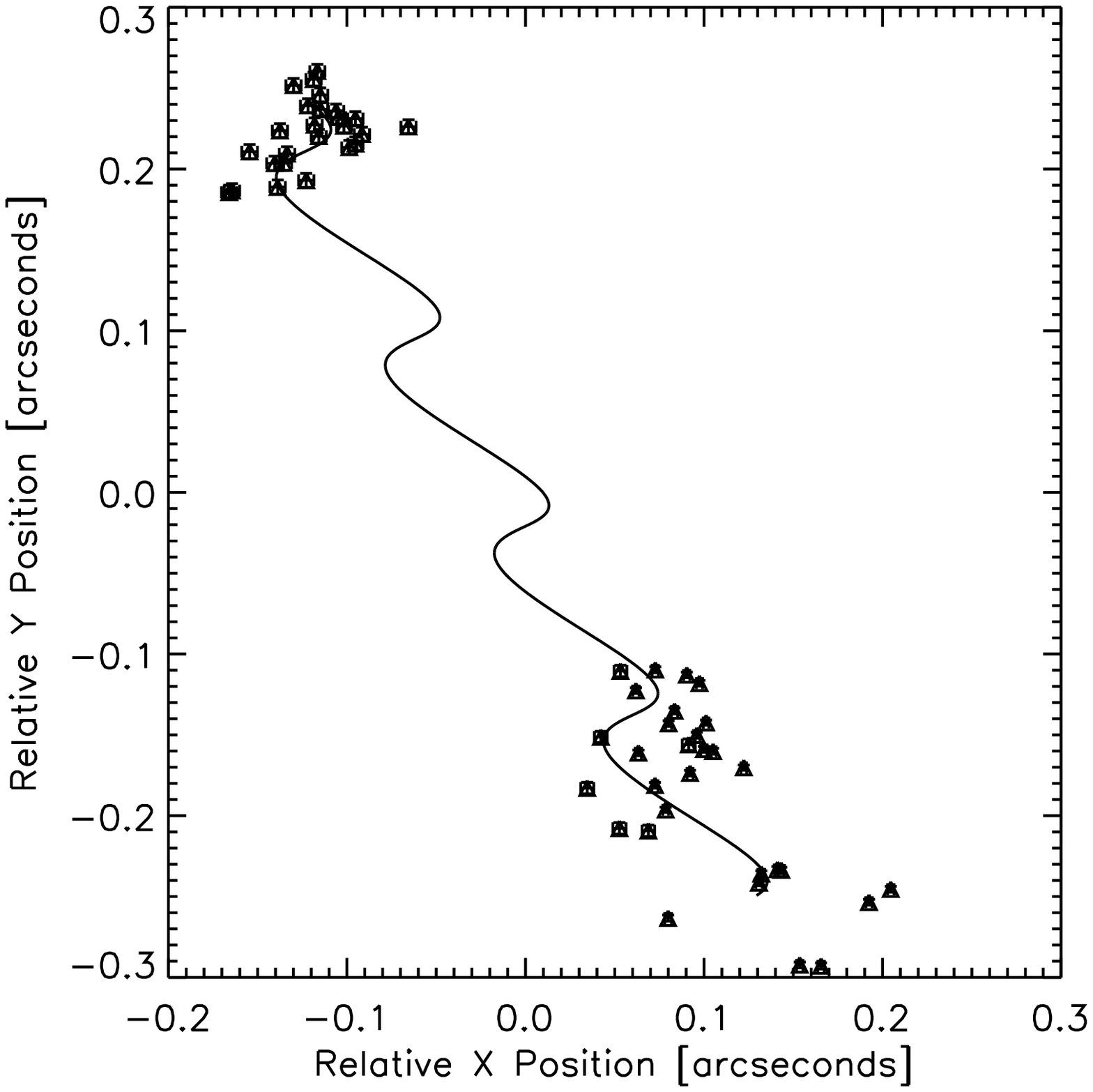}{0.25\textwidth}{2M1017+1308}
          \fig{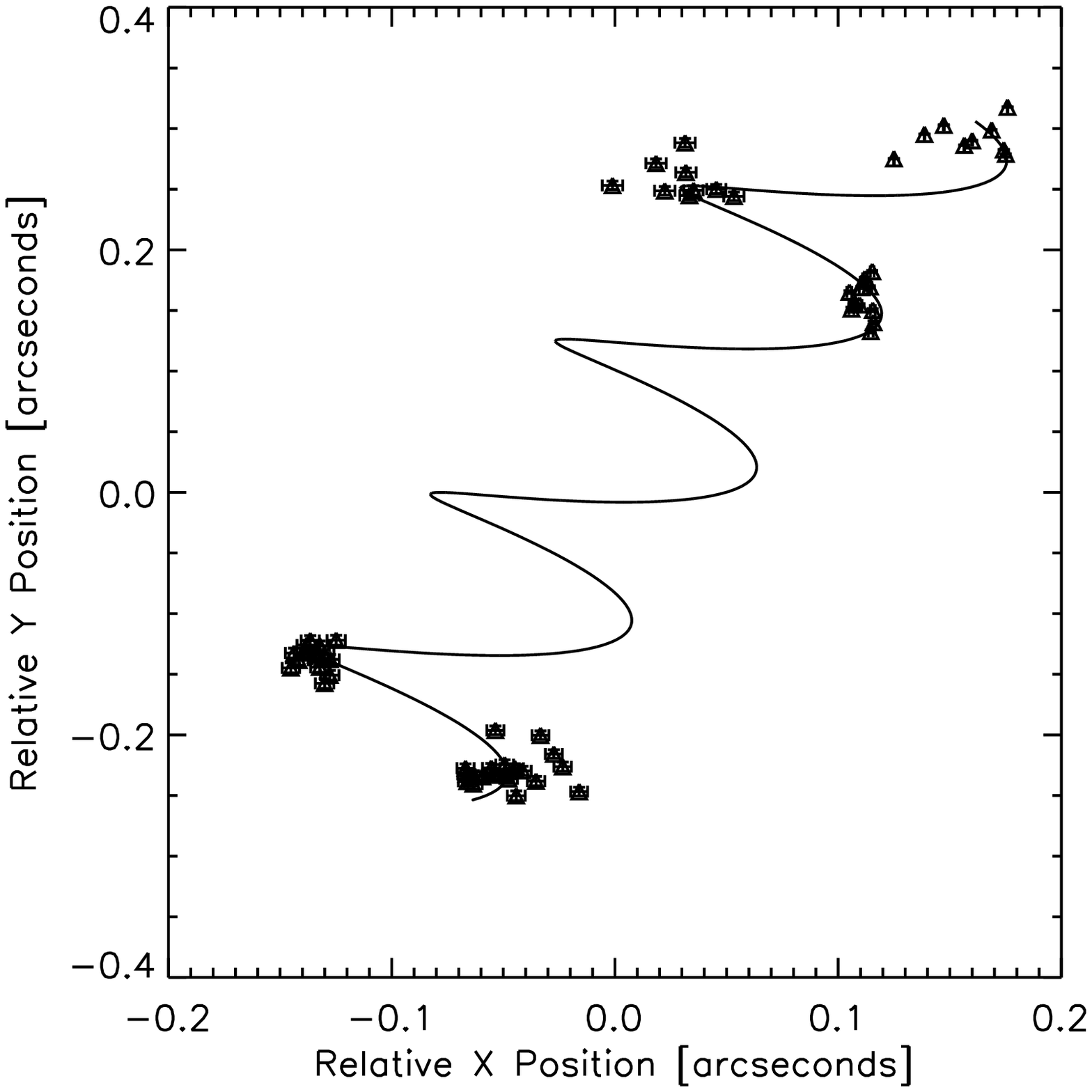}{0.25\textwidth}{2M1104+1959}
          \fig{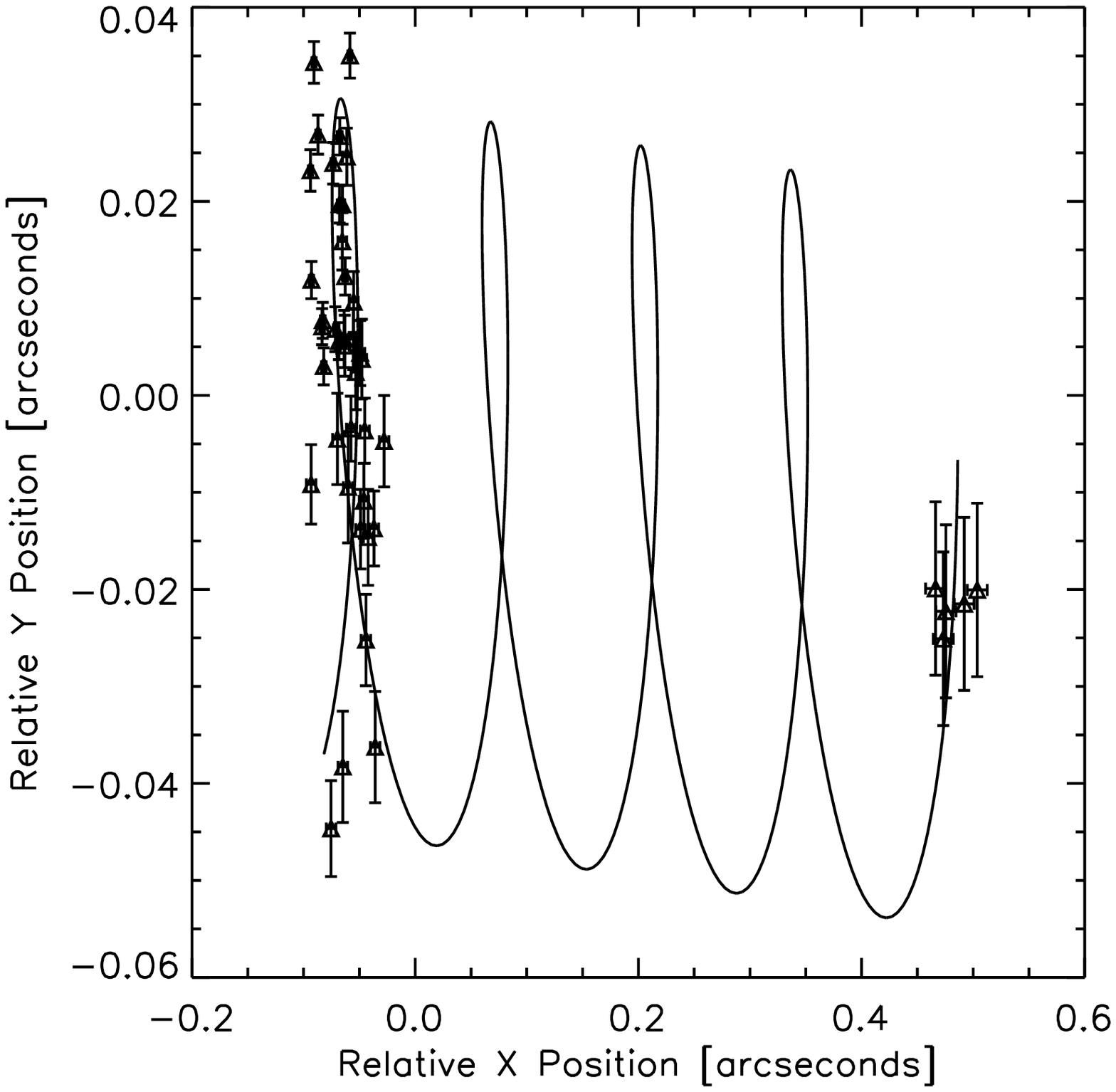}{0.25\textwidth}{2M1239+5515}
          \fig{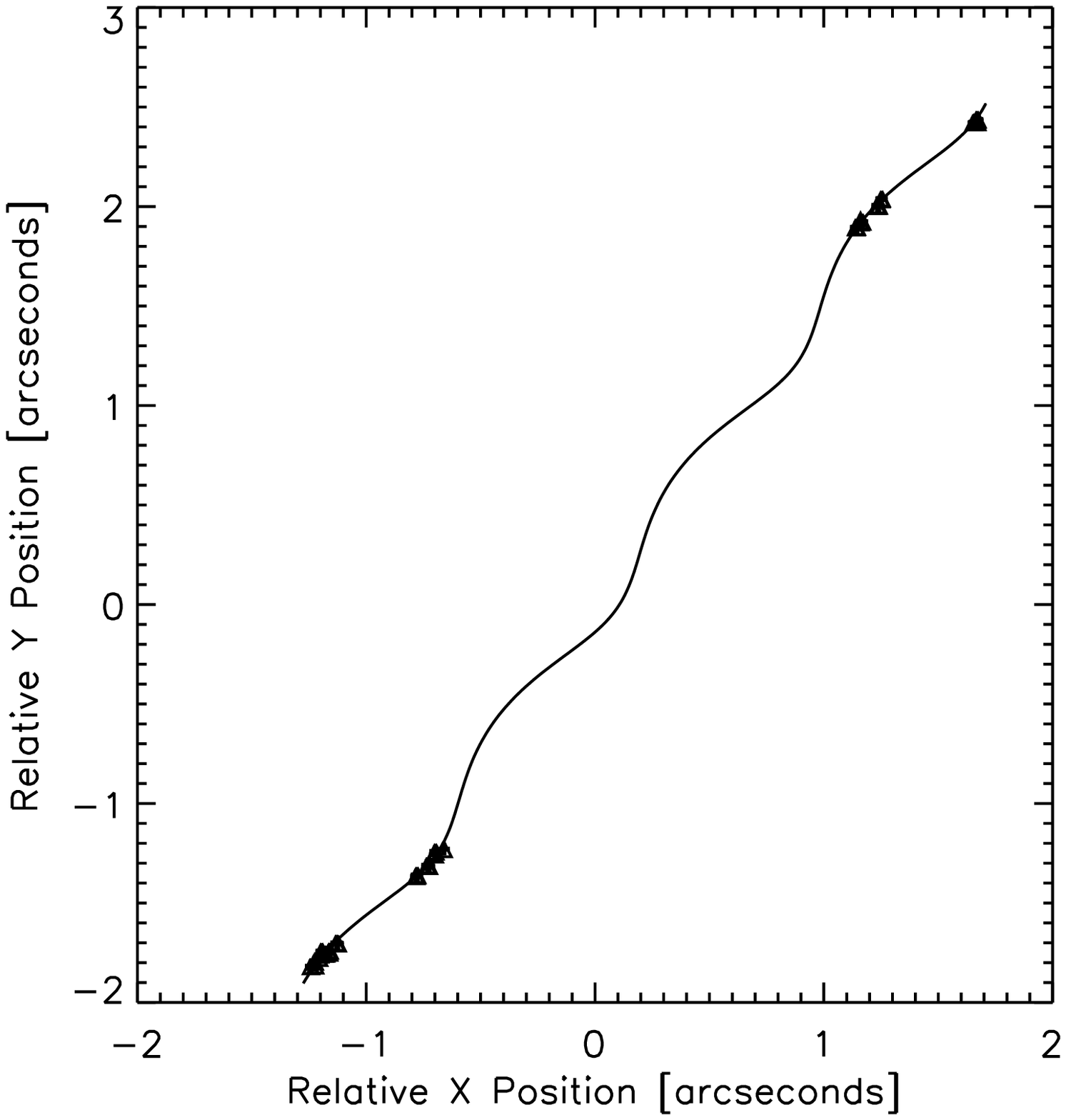}{0.25\textwidth}{2M1300+1912}
          }
\gridline{\fig{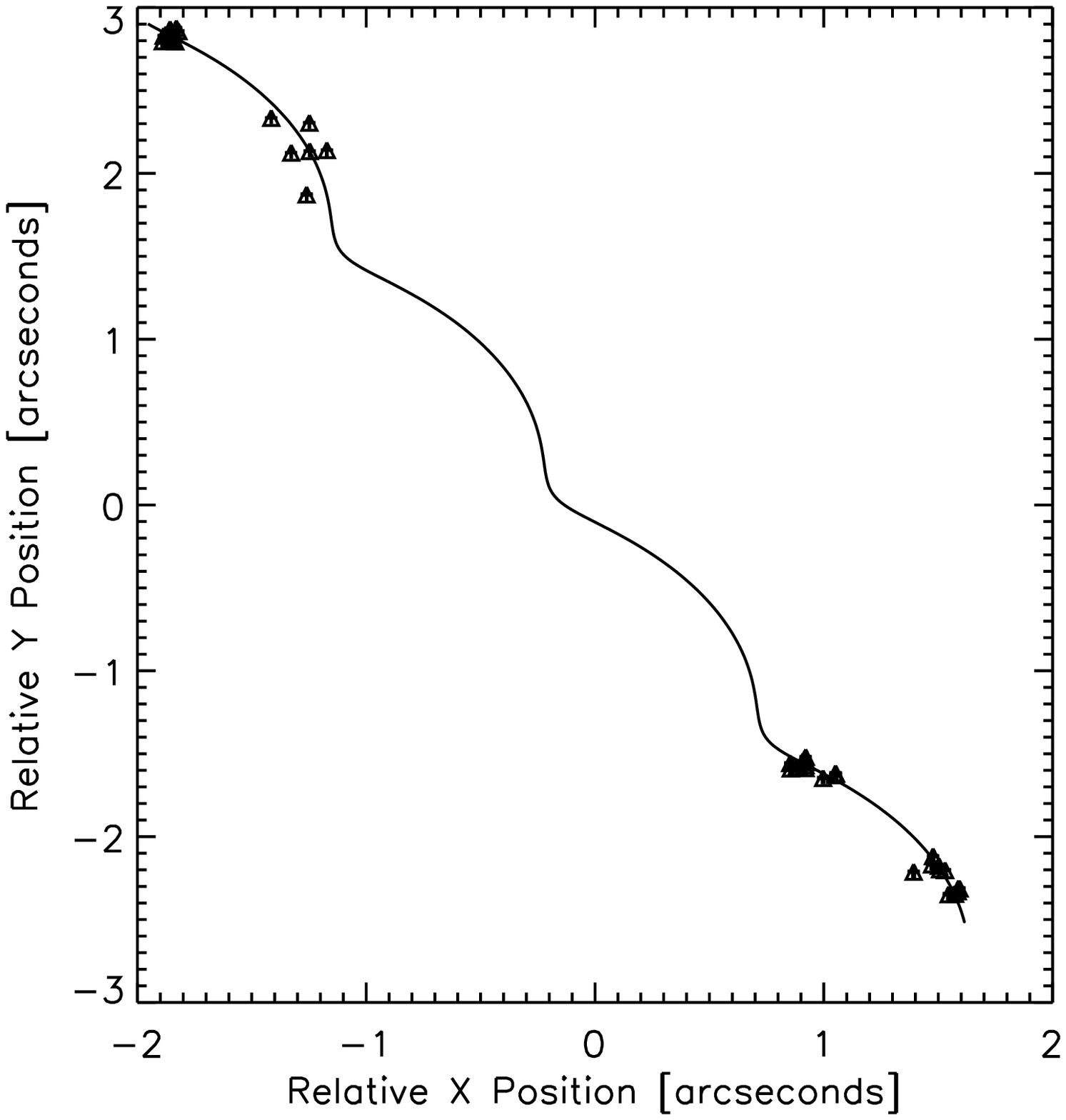}{0.25\textwidth}{2M1515+4847}
          \fig{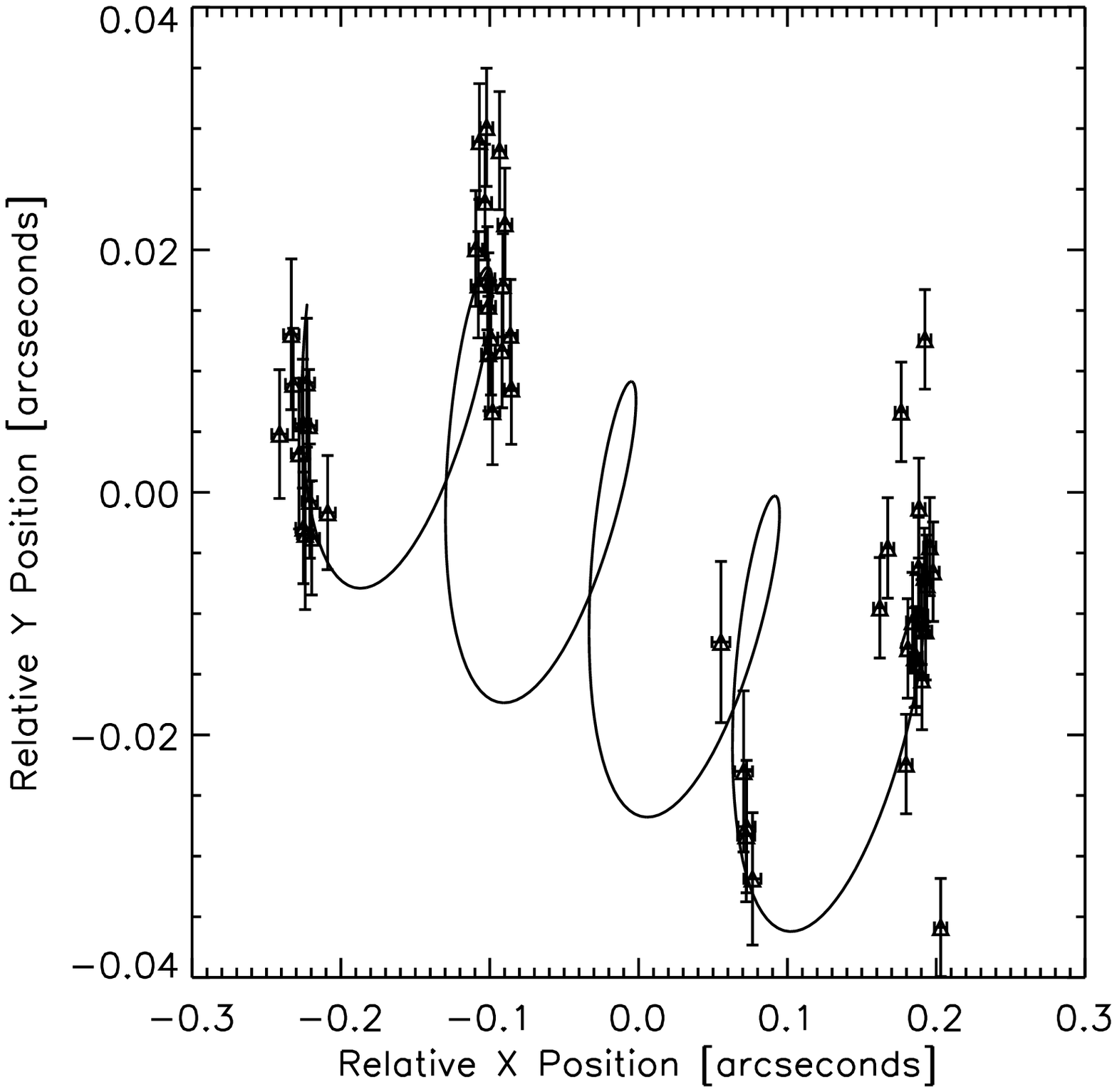}{0.25\textwidth}{2M2028+0052} 
         }
\caption{Predicted sky motion and observations of the LT parallax targets over the observational program.}
\label{elis}
\end{figure*}


\end{document}